\documentclass[preprint]{JHEP3} 

\JHEPspecialurl{http://jhep.sissa.it/JOURNAL/JHEP3.tar.gz}

\usepackage{epsfig,multicol,bbm}
\usepackage{axodraw}
\newcommand\fverbdo{\egroup\medskip\noindent%
            \fbox{\unhbox\fverbbox}\ }
\newcommand\fverbit{\egroup\item[\fbox{\unhbox\fverbbox}]}
\newbox\fverbbox

\title{Feynman rules for the rational part of the
Electroweak 1-loop amplitudes in the $R_{\xi}$ gauge and in the Unitary gauge.}

\author{}

\author{M.V. Garzelli\\
       Departamento de F\'{i}sica Te\'orica y del Cosmos y CAFPE
       Universidad de Granada, E-18071 Granada, Spain and \\
       INFN Milano, I-20133 Milano, Italy. \\
       E-mail: \email{garzelli@to.infn.it}}

\author{I. Malamos\\
        Department of Theoretical High Energy Physics,
        Institute for Mathematics, Astrophysics and Particle Physics, 
        Radboud Universiteit Nijmegen,  
        6525 AJ Nijmegen, the Netherlands. \\ 
        E-mail: \email{J.Malamos@science.ru.nl}}

\author{R. Pittau\\
       Departamento de F\'{i}sica Te\'orica y del Cosmos y CAFPE
       Universidad de Granada, E-18071 Granada, Spain.\\
       E-mail: \email{pittau@ugr.es}}

\abstract{
We present the complete set of Feynman rules producing the rational terms of kind ${\rm R_2}$ needed to perform any 1-loop calculation in the Electroweak Standard Model.
Our formulae are given both in the $R_{\xi}$ gauge and in the Unitary gauge, therefore 
completing the results in the 't Hooft-Feynman gauge already presented in a 
previous publication.  

As a consistency check, we verified, in the case of the process $H \to \gamma \gamma$ and in a few other physical cases, the independence of the total Rational Part (${\rm R_1+R_2}$) on the chosen gauge.
In addition, we explicitly checked the equivalence of the limits $\xi \to \infty$ after or before the loop momentum 
integration in the definition of the Unitary gauge at 1-loop.}


\begin{document}

\newcounter{im}
\setcounter{im}{0}
\newcommand{\exampleSp}{\stepcounter{im}\includegraphics[scale=0.9]{SpinorExamples_\arabic{im}.eps}}
\newcommand{\myindex}[1]{\label{com:#1}\index{{\tt #1} & pageref{com:#1}}}
\renewcommand{\topfraction}{1.0}
\renewcommand{\bottomfraction}{1.0}
\renewcommand{\textfraction}{0.0}
\newcommand{\nn}{\nonumber \\}
\newcommand{\eqn}[1]{eq.~\ref{eq:#1}}
\newcommand{\be}{\begin{equation}}
\newcommand{\ee}{\end{equation}}
\newcommand{\ba}{\begin{array}}
\newcommand{\ea}{\end{array}}
\newcommand{\bea}{\begin{eqnarray}}
\newcommand{\eea}{\end{eqnarray}}
\newcommand{\bqa}{\begin{eqnarray}}
\newcommand{\eqa}{\end{eqnarray}}
\newcommand{\nl}{\nonumber \\}
\def\db#1{\bar D_{#1}}
\def\zb#1{\bar Z_{#1}}
\def\d#1{D_{#1}}
\def\tld#1{\tilde {#1}}
\def\slh#1{\rlap / {#1}}
\def\eqn#1{eq.~(\ref{#1})}
\def\eqns#1#2{Eqs.~(\ref{#1}) and~(\ref{#2})}
\def\eqnss#1#2{Eqs.~(\ref{#1})-(\ref{#2})}
\def\fig#1{Fig.~{\ref{#1}}}
\def\figs#1#2{Figs.~\ref{#1} and~\ref{#2}}
\def\sec#1{Section~{\ref{#1}}}
\def\app#1{Appendix~\ref{#1}}
\def\tab#1{Table~\ref{#1}}
\def\cg{c_\Gamma}
\newcommand{\bfig}{\begin{center}\begin{picture}}
\newcommand{\efig}[1]{\end{picture}\\{\small #1}\end{center}}
\newcommand{\flin}[2]{\ArrowLine(#1)(#2)}
\newcommand{\ghlin}[2]{\DashArrowLine(#1)(#2){5}}
\newcommand{\wlin}[2]{\DashLine(#1)(#2){2.5}}
\newcommand{\zlin}[2]{\DashLine(#1)(#2){5}}
\newcommand{\glin}[3]{\Photon(#1)(#2){2}{#3}}
\newcommand{\gluon}[3]{\Gluon(#1)(#2){5}{#3}}
\newcommand{\lin}[2]{\Line(#1)(#2)}
\newcommand{\sof}{\SetOffset}



\section{Introduction}
The complete automation of the 1-loop calculations is nowadays a 
feasible task~\cite{nlopapers}. 
The advent of the OPP reduction method~\cite{opp},
together with the concept of multiple cuts~\cite{britto}, 
allowed to revitalize Unitarity~\cite{unitarity} based Techniques,
such as Generalized Unitarity (GU)~\cite{genunit}, by reducing 
the computation of 1-loop amplitudes to a problem with the same conceptual 
complexity of a tree level calculation, resulting in achievements that 
were inconceivable only a few years ago~\cite{proof}.
As a matter of principle, any program capable of producing tree level results
can be transformed nowadays into a NLO calculator by either {\em cutting} 
the 1-loop diagrams, in the OPP method, or by {\em gluing} tree level
structures in the GU approach.

Both OPP and GU, when applied in 4 dimensions, allow the extraction of 
the Cut Constructible (CC) part of the amplitude, while a 
left over piece, the rational part ${\rm R}$, needs to be derived separately.
In the Generalized Unitarity approaches, this is achieved by computing
the amplitude in different numbers of space-time dimensions, or 
via bootstrapping techniques~\cite{boot}, while, 
in the OPP approach, ${\rm R}$ is split in 2 pieces ${\rm R= R_1+R_2}$. 
The first piece, ${\rm R_1}$, is derivable in the same framework used 
to reconstruct the CC part of the amplitude, while ${\rm R_2}$ is computable 
through a special set of Feynman rules for the theory at hand~\cite{rational}, 
to be used in a tree level-like computation. 

The OPP treatment of ${\rm R}$ in its present formulation
has one advantage and one drawback. 
The advantage is that no calculation in dimensions other than four
is needed, avoiding the use of 6 and 8 dimensional
explicit representations of the external particle wave functions.
The drawback is that, for each theory that needs to be studied, 
a different special set
of Feynman Rules has to be explicitly computed once for all. 
On the other hand, in the OPP framework, the speed for computing 
the Rational part is very high, so that we prefer it.

The full set of ${\rm R_2}$ Feynman rules has been already derived for QED 
in~\cite{rational}, for QCD in~\cite{qcdrational}, and, for the 
Standard Model (SM) of the Electroweak (EW) interactions in the 
't Hooft-Feynman gauge in~\cite{ewrational}.
It is the main aim of the present paper to present the ${\rm R_2}$ Feynman 
rules for the Electroweak Standard Model in a general renormalizable
$R_{\xi}$ gauge and in the Unitary gauge. On the one hand,
this completes the theoretical picture on $R_2$ and, 
on the other hand, it allows tree level packages based on gauges 
other that the 't Hooft-Feynman one to be transformed 
into 1-loop calculators with the help of the mentioned 
OPP or GU techniques.
In addition, the use of a general renormalizable $R_{\xi}$ gauge, can be used
to verify the correctness and the numerical stability 
of the 1-loop predictions by studying the invariance 
of the results under a change in the numerical value of $\xi$.

The outline of the paper is as follows. In section~\ref{sec:2} we recall the origin of ${\rm R_2}$.
In section~\ref{sec:3} we fix our notation and our  calculational framework.
Section~\ref{sec:4} contains the complete
list of all possible special ${\rm R_2}$ EW SM vertices in  the $R_{\xi}$ gauge and in the Unitary gauge.
Finally, section~\ref{checks} describes the tests we performed on our formulae.

\section{Theory of ${\rm R_2}$  \label{sec:2}}
The presence of a rational part ${\rm R}$ in a generic 1-loop amplitude is due to the regularization
procedure needed before carrying out the calculation.
In dimensional regularization, one computes the 1-loop integrals in  
$n=~4+\epsilon$ dimensions, so that a generic $m$-point one-loop (sub-) amplitude reads
\bqa
\label{eq:1}
{\cal A}=   \frac{1}{(2 \pi)^4} \int d^n \bar q \frac{\bar N(\bar q)}{\db{0}\db{1}\cdots \db{m-1}}\,,~~~
\db{i} = ({\bar q} + p_i)^2-m_i^2\,,~~~\bar q = q + \tld{q}\,,
\eqa
where ${\bar q}$ is the integration momentum, a bar denotes objects living
in $n$ dimensions and a tilde represents $\epsilon$-dimensional quantities.
Notice that the external momenta $p_i$ are always kept in 4 dimensions.

The numerator function $\bar{N}(\bar q)$ can be 
split into a $4$-dimensional plus an $\epsilon$-dimensional part
\bqa
\label{eq:split}
\bar{N}(\bar q) = N(q) + \tld{N}(\tld{q}^2,q,\epsilon)\,.
\eqa
$N(q)$ brings information on the CC part of the amplitude (and within the OPP
framework can also be used to compute a part of the rational piece called ${\rm R_1}$), while
$\tld{N}(\tld{q}^2,q,\epsilon)$ gives rise to a second piece of the rational part called ${\rm R_2}$, defined as
\bqa
\label{eqr2}
{\rm R_2} \equiv  \frac{1}{(2 \pi)^4}\int d^n\,\bar q
\frac{\tld{N}(\tld{q}^2,q,\epsilon)}{\db{0}\db{1}\cdots \db{m-1}} \,.
\eqa
Due to possible ambiguities when passing from $N(q)$ to $\bar{N}(\bar q)$, the actual form of
$\tld{N}(\tld{q}^2,q,\epsilon)$ can only be read, to the best of our knowledge, starting from the original theory in
$n$ dimensions. In the OPP framework, that is achieved by computing 
analytically tree-level like Feynman rules, by splitting the Feynman diagrams according to the following 
three rules 
\bqa
\label{qandg}
\bar q_{\bar \mu}                 &=& q_\mu + \tld{q}_{\tld{\mu}}\,, \nl
\bar \gamma_{\bar \mu} &=&  \gamma_{\mu}+ \tld{\gamma}_{\tld{\mu}}\,,\nl
 \bar g^{\bar \mu \bar \nu}  &=&  g^{\mu \nu}+  \tld{g}^{\tld{\mu} \tld{\nu}}\,.
\eqa
Effective vertices are then generated by calculating the ${\rm R_2}$ parts coming from
all possible one-particle irreducible Green functions of the theory at hand, 
up to four external legs.
The fact that four external legs are enough to account for ${\rm R_2}$ 
is guaranteed by the ultraviolet nature of the rational terms, 
proved in~\cite{directcomp1}
\footnote{In GU approaches, the entire calculation is instead performed in $n$ dimensions, at the price of
introducing, as already mentioned in the Introduction, explicit 6 and 8-dimensional polarization vectors for the particles glued together
to form the loop amplitude.}.
Some freedom is however left in the choice of the regularization procedure, so that, instead of
Eq.~\ref{eqr2}, one could also use the definition
\bqa
\label{eqr2fdh}
{\rm R_2} \Bigl |_{FDH} =  \frac{1}{(2 \pi)^4}\int d^n\,\bar q
\frac{\tld{N}(\tld{q}^2,q,\epsilon= 0)}{\db{0}\db{1}\cdots \db{m-1}} \,,
\eqa
provided the same prescription is used in all parts of the calculation.
The choice in Eq.~\ref{eqr2fdh} corresponds to the so called Four Dimensional Helicity scheme~\cite{fdhqcd} (FDH).
In such a scheme, when using dimensional regularization, the only object 
to be continued in $n$ dimensions is 
\bqa
\label{eq:repl}
q^2 \to q^2 + \tld{q}^2\,,
\eqa
and it would be nice if one could be able to use this information 
to have access to ${\rm R_2}$ starting uniquely from the theory in 4 dimensions.
Unfortunately, the replacement in Eq.~\ref{eq:repl} is still too ambiguous, in the sense that different ways
of writing $N(q)$ may lead to different $n$-dimensional continuations, as already observed in~\cite{ewrational}, so that no better solution can be found, at present, than relying on the original 
$n$-dimensional theory.
It is worth mentioning that only the combination ${\rm R}= {\rm R_1}+{\rm R_2}$ is gauge invariant, not, in general,
${\rm R_1}$ or ${\rm R_2}$ separately. In this respect, the {\em right} analytical continuation from 
$N(q)$ to $\bar{N}(\bar q)$ by means of Eq.~\ref{eq:repl}, is that one that preserves all the Ward Identities of the theory.

In the following sections, we present the result of the explicit calculation we performed of all possible 2, 3 and 4-point
effective vertices in the Electroweak Standard Model in a general $R_{\xi}$ gauge
and in the Unitary gauge.

\section{Notations and Feynman rules\label{sec:3}}
The vector boson fields (generically symbolized by $V$) 
are denoted by $A$, $Z$, $W^\pm$. The physical scalar Higgs field 
is written as $H$ while $\chi$ and $\phi^\pm$ denote the neutral and the charged
scalar goldstone bosons, respectively. All scalar fields are 
generically symbolized by $S$.
We work in the 1-fermion-family approximation, with lepton and quark doublets
given by 
\bqa
\label{doublets}
\left(
\begin{tabular}{l}
$\nu_l$ \\
$l$ 
\end{tabular}
\right)
\hspace{1cm}{\rm and}\hspace{1cm}
\left(
\begin{tabular}{l}
$u$ \\
$d$ 
\end{tabular}
\right)\,.
\eqa
Fermions are generically symbolized by $f$, and the charge,
 the third isospin component and the mass of a fermion
by $Q_f$, $I_{3f}$ and $m_f$, respectively.

The sine and cosine of the Weinberg angle, the $W$ and the $Z$ mass 
are denoted by $c_w$, $s_w$, $M_W$ and $M_Z$, respectively.
Following reference~\cite{denner}, we introduce the two quantities 
$V_{ud}$ and $V^{\dagger}_{du}$ in the coupling of the $W$ boson with 
the quark doublet of Eq.~\ref{doublets}.
This allows one to keep track of the CKM matrix and to easily 
generalize the results to the 3-families case.
Finally, we use projector operators denoted  by 
$\Omega^\pm = \frac{1 \pm \gamma_5}{2}$.

The set of Feynman rules we use for our calculation 
is that one given in~\cite{denner}, with some modifications due to the fact that
the expressions in that paper refer to the 't Hooft-Feynman gauge, while
we want to work in the $R_\xi$ gauge. 
In the computation of ${\rm R_2}$, the ghost fields never enter, so that, 
in order to pass from the  the 't Hooft-Feynman gauge to the $R_\xi$ one, 
we just need to modify the propagators of the scalar goldstone bosons and
of the vector bosons as follows
\begin{center}
  \begin{picture}(300,100)
   \SetScale{0.5}
    \SetWidth{0.5}
    \SetColor{Black}
    \SetOffset(0,105)
    \Text(64,-17)[]{{\Black{$p$}}}
    \LongArrow(105,-48)(140,-48)  
    \DashLine(56,-62)(188,-62){8}
    \Text(104,-31)[]{$S$} 
    \Text(18,-31)[]{$S$} 
    \Text(120,-31)[l]{$\displaystyle = \frac{i}{p^2-\xi M_S^2} $}
    \SetOffset(0,50)
    \Text(64,-17)[]{{\Black{$p$}}}
    \LongArrow(105,-48)(140,-48)  
    \Photon(56,-62)(188,-62){5.5}{8}
    \Text(104,-31)[]{$V_\beta$} 
    \Text(18,-31)[]{$V_\alpha$} 
    \Text(120,-31)[l]{$\displaystyle = 
 \frac{-i}{p^2-M_V^2} \left(g_{\alpha\beta}
  -(1-\xi)\frac{p_\alpha p_\beta}{p^2-\xi M_V^2}\right)\,.
$}
  \end{picture}
\end{center}
To compute our results in the Unitary gauge, we simply take the
limit $\xi \to \infty$ in the above propagators {\em before integrating}
over the loop momentum
\footnote{See section~\ref{checks} for more discussions on this issue.}.
Then the unphysical scalar particles decouple and the massive 
gauge boson propagators become
\bqa
 \label{eqmasspro}
 \frac{-i}{p^2-M_V^2} \left(g_{\alpha\beta}-\frac{p_\alpha p_\beta}{M_V^2}\right)\,,
\eqa
while for the photon we use
\bqa
 \frac{-i}{p^2} g_{\alpha\beta}.
\eqa
Notice that the choice in Eq.~\ref{eqmasspro} is mandatory in the framework 
of the OPP method. 
In fact, taking the limit $\xi \to \infty$ {\em after} integration over
the loop momentum would imply a nonviable numerical 
cancellation between ${\rm R_1}$ and ${\rm R_2}$, since the two parts are treated separately. 

A last comment is in order with respect to our 
treatment of $\gamma_5$ in vertices 
containing fermionic lines. When computing all contributing Feynman diagrams, 
we pick up a ``special'' vertex  in the loop and anticommute all $\gamma_5$'s 
to reach it before performing the $n$-dimensional algebra, and,  when a trace 
is present, we start reading it from this vertex. This treatment
produces, in general, a term proportional to the totally antisymmetric $\epsilon$ 
tensor, whose coefficient may be different 
depending on the choice of the ``special'' 
vertex. However, when summing over all quantum numbers of each fermionic 
family, we checked that all contributions proportional to $\epsilon$ cancel.

\section{Results \label{sec:4}}
In this section, we present our results.
We omit, in this paper, the gauge invariant contributions 
coming from fermion loops, because they can be recovered with the help 
of the formulae we already worked out in the case of the 't Hooft-Feynman 
gauge in~\cite{ewrational}.
In fact, the fermion loop part can be easily separated from the rest since it 
always involves a sum $\sum_{i}$ over fermions or fermion families.
A parameter $\lambda_{HV}$ is introduced in our formulae such that 
$\lambda_{HV}= 1$ corresponds to the 't Hooft-Veltman scheme
and $\lambda_{HV}= 0$ to the FDH scheme of eq.~\ref{eqr2fdh}.

\begin{figure}
\begin{center}
\begin{tabular}{l}
  \begin{picture}(200,100)
    \SetOffset(-10,45)
    \SetWidth{0.5}
    \SetScale{0.5}
    \SetColor{Black}
    \DashLine(57,-36)(161,54){8}
    \DashLine(57,54)(161,-36){8}
    \Vertex(110,9){6.0}
    \LongArrow(60,-17)(85,3)
    \Text(27.5,-7.5)[tr]{{\Black{$p_1$}}}
    \LongArrow(60,33)(85,13)
    \Text(26,20)[tr]{{\Black{$p_2$}}}
    \LongArrow(145,55)(120,33) 
    \Text(55,25)[lb]{{\Black{$p_3$}}}
    \LongArrow(145,-37)(120,-14) 
    \Text(55,-15)[lt]{{\Black{$p_4$}}}
    \Text(22,-18)[tr]{$S_1$}
    \Text(22, 27)[br]{$S_2$}
    \Text(90, 27)[bl]{$S_3$}
    \Text(90,-18)[tl]{$S_4$}
    \Text(115,5)[l]{$= ~{\rm Vert}(S_1,S_2,S_3,S_4)$}
  \end{picture} \\
  \begin{picture}(200,100)
    \SetOffset(-10,45)
    \SetWidth{0.5}
    \SetScale{0.5}
    \SetColor{Black}
    \DashLine(57,-36)(110,9){8}
    \Photon(110,9)(161,54){5.5}{4}
    \DashLine(57,54)(110,9){8}
    \Photon(110,9)(161,-36){5.5}{4}
    \Vertex(110,9){6.0}
    \LongArrow(60,-17)(85,3)
    \Text(27.5,-7.5)[tr]{{\Black{$p_1$}}}
    \LongArrow(60,33)(85,13)
    \Text(26,20)[tr]{{\Black{$p_2$}}}
    \LongArrow(145,55)(120,33) 
    \Text(55,25)[lb]{{\Black{$p_3$}}}
    \LongArrow(145,-37)(120,-14) 
    \Text(55,-15)[lt]{{\Black{$p_4$}}}
    \Text(22,-18)[tr]{$S_1$}
    \Text(22, 27)[br]{$S_2$}
    \Text(90, 27)[bl]{$V_{1\gamma}$}
    \Text(90,-18)[tl]{$V_{2\delta}$}
    \Text(115,5)[l]{$= ~{\rm Vert}(S_1,S_2,V_1,V_2)$}
  \end{picture}\\
  \begin{picture}(200,100)
    \SetOffset(-10,45)
    \SetWidth{0.5}
    \SetScale{0.5}
    \SetColor{Black}
    \Photon(57,-36)(161,54){5.5}{8}
    \Photon(57,54)(161,-36){5.5}{8}
    \Vertex(110,9){6.0}
    \LongArrow(60,-17)(85,3)
    \Text(27.5,-7.5)[tr]{{\Black{$p_1$}}}
    \LongArrow(60,33)(85,13)
    \Text(26,20)[tr]{{\Black{$p_2$}}}
    \LongArrow(145,55)(120,33) 
    \Text(55,25)[lb]{{\Black{$p_3$}}}
    \LongArrow(145,-37)(120,-14) 
    \Text(55,-15)[lt]{{\Black{$p_4$}}}
    \Text(22,-18)[tr]{$V_{1\alpha}$}
    \Text(22, 27)[br]{$V_{2\beta}$}
    \Text(90, 27)[bl]{$V_{3\gamma}$}
    \Text(90,-18)[tl]{$V_{4\delta}$}
    \Text(115,5)[l]{$= ~{\rm Vert}(V_1,V_2,V_3,V_4)$}
  \end{picture}
\end{tabular}
\end{center}
\caption{\label{fig:3} All possible non vanishing 4-point vertices.}
\end{figure}
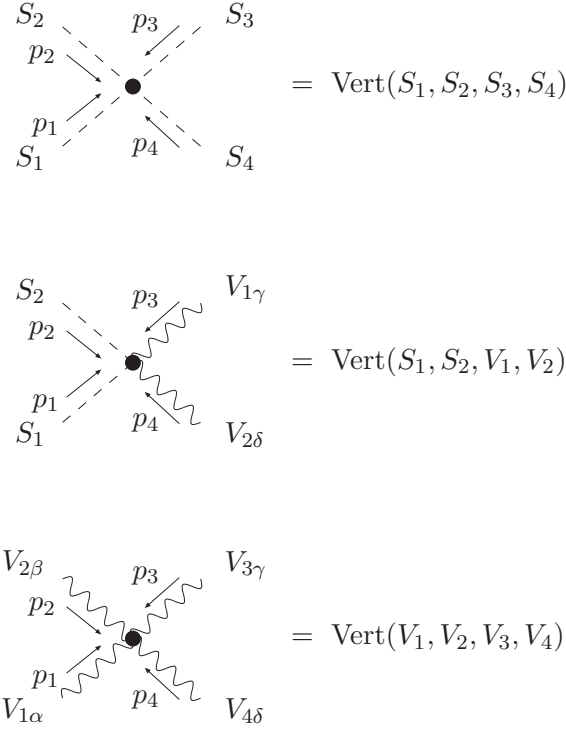

We explicitly write down, in this publication, all the formulae 
in the 2-point case, while, 
for the 3 the and 4-point vertices, we just classify the non vanishing ones.
In fact the expression we obtained are rather lengthy, and there is no point
in writing them down on paper. We rather provide the full set of results as  
FORM~\cite{form} files~\cite{formfiles}. The notation used in 
those files closely follows that one introduced in the previous section.
In Fig.~\ref{fig:1}-\ref{fig:3} we present the generic non vanishing 
2-point, 3-point and 4-point vertices that appear in our calculation,
that also serve to further fix our notations.

\begin{figure}
\begin{center}
\begin{tabular}{l}
  \begin{picture}(200,50)
   \SetScale{0.5}
    \SetWidth{0.5}
    \SetColor{Black}
    \SetOffset(0,55)
    \Vertex(142,-62){6.0}
    \Text(50,-17)[]{{\Black{$p_1$}}}
    \LongArrow(75,-48)(110,-48)  
    \DashLine(56,-62)(142,-62){8}
    \DashLine(142,-62)(228,-62){8}
    \Text(124,-31)[]{$S_2$} 
    \Text(18,-31)[]{$S_1$} 
    \Text(138,-31)[l]{$= ~{\rm Vert}(S_1,S_2)$}
    \Text(-90,-31)[]{$(a)$}
  \end{picture} \\
  \begin{picture}(200,50) 
    \SetOffset(0,42)
    \SetScale{0.5}
    \SetWidth{0.5}
    \SetColor{Black}
       \Photon(56,-34)(133,-34){5.5}{3.6}
     \DashLine(135,-34)(228,-34){8}
    \Vertex(142,-34){6.0}
    \Text(50,-3)[]{{\Black{$p_1$}}}
    \LongArrow(75,-20)(110,-20)  
    \Text(18,-17)[]{$V_{\alpha}$}
    \Text(124,-17)[]{$S$}
    \SetOffset(0,55)
    \Text(138,-31)[l]{$= ~{\rm Vert}(V,S)$}
    \Text(-90,-31)[]{$(b)$}
  \end{picture} \\
  \begin{picture}(200,50) 
    \SetOffset(0,42)
    \SetScale{0.5}
    \SetWidth{0.5}
    \SetColor{Black}
    \Photon(56,-34)(133,-34){5.5}{3.6}
    \Vertex(142,-34){6.0}
    \Photon(135,-34)(228,-34){5.5}{4}
    \Text(50,-3)[]{{\Black{$p_1$}}}
    \LongArrow(75,-20)(110,-20)  
    \Text(18,-17)[]{$V_{1\alpha}$}
    \Text(124,-17)[]{$V_{2\beta}$}
    \SetOffset(0,55)
    \Text(138,-31)[l]{$= ~{\rm Vert}(V_1,V_2)$}
    \Text(-90,-31)[]{$(c)$}
  \end{picture}  \\
  \begin{picture}(200,50) 
   \SetOffset(0,55)
    \SetWidth{0.5}
    \SetScale{0.5}
    \SetColor{Black}
    \ArrowLine(56,-62)(138,-62)
    \ArrowLine(141,-62)(228,-62)
    \Vertex(142,-63){6.0}
    \Text(50,-17)[]{{\Black{$p_1$}}}
    \LongArrow(75,-48)(110,-48)  
    \DashLine(56,-62)(142,-62){8}
    \Text(18,-31)[]{$f_1$}
    \Text(124,-31)[]{$\bar f_2$}
    \Text(138,-31)[l]{$= ~{\rm Vert}(f_1,f_2)$}
    \Text(-90,-31)[]{$(d)$}
  \end{picture}  
\end{tabular}
\end{center}
\caption{\label{fig:1} All possible 2-point vertices.}
\end{figure}

\begin{figure}
\begin{center}
\begin{tabular}{l}
  \begin{picture}(210,70) 
    \SetOffset(-10,25)
    \SetWidth{0.5}
    \SetScale{0.5}
    \SetColor{Black}
    \Vertex(146,15){6.0}
    \DashLine(72,15)(146,15){8}
    \DashLine(146,15)(200,57){8}
    \DashLine(200,-27)(146,15){8}
    \LongArrow(85,28)(120,28)
    \Text(52,18)[b]{{\Black{$p_1$}}}
    \LongArrow(180,62.3)(156,43) 
    \LongArrow(180,-31)(156,-12) 
    \Text(75,29)[lb]{{\Black{$p_2$}}}
    \Text(75,-13)[lt]{{\Black{$p_3$}}}
    \Text(33,7.5)[r]{\Black{$S_1$}}
    \Text(107, 30)[l]{$S_2$}
    \Text(107,-15)[l]{$S_3$}
    \Text(125,7.5)[l]{$= ~{\rm Vert}(S_1,S_2,S_3)$}
   \end{picture} \\
  \begin{picture}(210,70) 
    \SetOffset(-10,25)
    \SetWidth{0.5}
    \SetScale{0.5}
    \SetColor{Black}
    \Vertex(152,15){6.0}
    \Photon(72,15)(146,15){5.5}{4}
    \DashLine(146,15)(200,57){8}
    \DashLine(200,-27)(146,15){8}
    \LongArrow(85,28)(120,28)
    \Text(52,18)[b]{{\Black{$p_1$}}}
    \LongArrow(180,62.3)(156,43) 
    \LongArrow(180,-31)(156,-12) 
    \Text(75,29)[lb]{{\Black{$p_2$}}}
    \Text(75,-13)[lt]{{\Black{$p_3$}}}
    \Text(33,7.5)[r]{\Black{$V_\alpha$}}
    \Text(107, 30)[l]{$S_1$}
    \Text(107,-15)[l]{$S_2$}
    \Text(125,7.5)[l]{$= ~{\rm Vert}(V,S_1,S_2)$}
   \end{picture} \\
  \begin{picture}(210,70) 
    \SetOffset(-10,25)
    \SetWidth{0.5}
    \SetScale{0.5}
    \SetColor{Black}
    \Vertex(146,15){6.0}
    \DashLine(72,15)(146,15){8}
    \Photon(146,15)(200,57){5.5}{4}
    \Photon(200,-27)(146,15){5.5}{4}
    \LongArrow(85,28)(120,28)
    \Text(52,18)[b]{{\Black{$p_1$}}}
    \LongArrow(180,62.3)(156,43) 
    \LongArrow(180,-31)(156,-12) 
    \Text(75,29)[lb]{{\Black{$p_2$}}}
    \Text(75,-13)[lt]{{\Black{$p_3$}}}
    \Text(33,7.5)[r]{\Black{$S$}}
    \Text(107, 30)[l]{$V_{1\beta}$}
    \Text(107,-15)[l]{$V_{2\gamma}$}
    \Text(125,7.5)[l]{$= ~{\rm Vert}(S,V_1,V_2)$}
   \end{picture} \\
  \begin{picture}(210,70) 
    \SetOffset(-10,25)
    \SetWidth{0.5}
    \SetScale{0.5}
    \SetColor{Black}
    \Vertex(146,15){6.0}
    \Photon(72,15)(146,15){5.5}{4}
    \Photon(146,15)(200,57){5.5}{4}
    \Photon(200,-27)(146,15){5.5}{4}
    \LongArrow(85,28)(120,28)
    \Text(52,18)[b]{{\Black{$p_1$}}}
    \LongArrow(180,62.3)(156,43) 
    \LongArrow(180,-31)(156,-12) 
    \Text(75,29)[lb]{{\Black{$p_2$}}}
    \Text(75,-13)[lt]{{\Black{$p_3$}}}
    \Text(33,7.5)[r]{\Black{$V_{1\alpha}$}}
    \Text(107, 30)[l]{$V_{2\beta}$}
    \Text(107,-15)[l]{$V_{3\gamma}$}
    \Text(125,7.5)[l]{$= ~{\rm Vert}(V_1,V_2,V_3)$}
\end{picture} \\
  \begin{picture}(210,70) 
   \SetOffset(-10,25)
    \SetWidth{0.5}
    \SetScale{0.5}
    \SetColor{Black}
    \Vertex(146,15){6.0}
    \DashLine(72,15)(144,15){8}
    \ArrowLine(146,15)(200,57)
    \ArrowLine(200,-27)(146,15)
    \LongArrow(85,28)(120,28)
    \Text(52,18)[b]{{\Black{$p_1$}}}
    \LongArrow(180,62.3)(156,43) 
    \LongArrow(180,-31)(156,-12) 
    \Text(75,29)[lb]{{\Black{$p_2$}}}
    \Text(75,-13)[lt]{{\Black{$p_3$}}}
\Text(33, 7.5)[r]{$S$}
\Text(102, 30)[l]{$\bar f_1$}
\Text(102,-15)[l]{$f_2$}
\Text(125,7.5)[l]{$= ~{\rm Vert}(S,f_1,f_2)$}
  \end{picture} \\
  \begin{picture}(210,70) 
    \SetOffset(-10,25)
    \SetWidth{0.5}
    \SetScale{0.5}
    \SetColor{Black}
    \Vertex(146,15){6.}
    \ArrowLine(146,15)(200,57)
    \ArrowLine(200,-27)(146,15)
    \Photon(72,15)(146,15){5.5}{4}
    \LongArrow(85,28)(120,28)
    \Text(52,18)[b]{{\Black{$p_1$}}}
    \LongArrow(180,62.3)(156,43) 
    \LongArrow(180,-31)(156,-12) 
    \Text(75,29)[lb]{{\Black{$p_2$}}}
    \Text(75,-13)[lt]{{\Black{$p_3$}}}
    \Text(30,7.5)[r]{\Black{$V_\alpha$}}
    \Text(102, 30)[l]{$\bar f_1$}
    \Text(102,-15)[l]{$f_2$}
    \Text(125,7.5)[l]{$= ~{\rm Vert}(V,f_1,f_2)$}  
  \end{picture}
\end{tabular}
\end{center}
\caption{\label{fig:2} All possible non vanishing 3-point vertices.}
\end{figure}

\subsection{The $R_\xi$ gauge}
\subsubsection{Bosonic contribution to the vertices with 2 legs}

\vspace{0.3cm}
\leftline{{\bf Scalar-Scalar effective vertices}}
\vspace{0.3cm}

\noindent The generic effective vertex is
\bqa
{\rm Vert}(S_1,S_2) = \frac{ie^2}{16 \pi^2 s_w^2} C
\eqa
with ${\rm Vert}(S_1,S_2)$ given in fig.~\ref{fig:1} $(a)$ and 
with the actual values of $S_1$, $S_2$ and $C$
\bqa
\label{eq:vert1}
%
%
HH~~:~~C & = & 
\frac{m_W^2}{4} \left(1+2\xi-\xi^2-12\lambda_{HV}\right)
  \left(1+\frac{1}{2 c_w^4} \right)
+ p_1^2\,\frac{9-11\xi}{24} \left(1+\frac{1}{2 c_w^2} \right) 
\nonumber \\
\nonumber \\
%
%
\chi\chi~~:~~C & = & 
 \frac{m_W^2}{24 c_w^4} \left(1+2\xi^2-12\lambda_{HV}\right)
+\frac{m_W^2}{12} \left(1-2\xi+7\xi^2-12\lambda_{HV}\right)
\nonumber \\
&&- \frac{m_H^2}{12 c_w^2} \left(1-\frac{5}{2}\xi \right)
+ p_1^2\,\frac{9-11\xi}{24} \left(1+\frac{1}{2 c_w^2} \right) 
\nonumber \\
\nonumber \\
%
%
\phi^-\phi^+ ~~:~~C & = &
 \frac{m_W^2}{24 c_w^4} \left(1+2\xi^2-12\lambda_{HV}\right)
+\frac{m_W^2}{2 c_w^2} \left(\xi-\frac{3}{2}\xi^2 \right)
\nonumber \\
&& +\frac{m_W^2}{12} \left(1-8\xi+16\xi^2-12\lambda_{HV}\right)
\nonumber \\
&& - \frac{m_H^2}{12} \left(1-\frac{5}{2}\xi \right)
+ p_1^2\,\frac{9-11\xi}{24} \left(1+\frac{1}{2 c_w^2} \right) 
\eqa

\vspace{0.3cm}
\leftline{{\bf Vector-Scalar effective vertices}}
\vspace{0.3cm}

\noindent The generic effective vertex is

\bqa
{\rm Vert}(V,S) = \frac{ie^2}{\pi^2}\, C\, p_{1\alpha}
\eqa
with ${\rm Vert}(V,S)$ given in fig.~\ref{fig:1} $(b)$ and 
with the actual values of $V$, $S$ and $C$
\bqa
\label{eq:vert2}
%
%
W^- \phi^+~~:~~C & = &
-(1-\xi) \frac{M_W}{192 c_w^2 s_w^2}
\nonumber \\ 
\nonumber \\ 
%
%
W^+\phi^-~~:~~C & = &
(1-\xi) \frac{M_W}{c_w^2 s_w^2}
\nonumber \\ 
\nonumber \\ 
%
%
Z \chi ~~:~~C & = &
(1-\xi) \frac{i\,M_Z}{192 c_w^2 s_w^2}\left(1+2c_w^2 s_w^2 \right)
\nonumber \\ 
\nonumber \\ 
%
%
A \chi ~~:~~C & = &
(1-\xi) \frac{i\,M_Z}{96}\frac{c_w}{s_w}
\eqa
Notice that all these vertices vanish in the 't Hooft-Feynman gauge 
($\xi = 1$).

\vspace{0.3cm}
\leftline{{\bf Vector-Vector effective vertices}}
\vspace{0.3cm}

\noindent The generic effective vertex is
\bqa
{\rm Vert}(V_1,V_2) = \frac{ie^2}{8 \pi^2} 
\left(C_1 \,p_{1\alpha} p_{1\beta} + C_2 \,g_{\alpha \beta}\right)
\eqa
with ${\rm Vert}(V_1,V_2)$ given in fig.~\ref{fig:1} $(c)$ and 
with the actual values of $V_1$, $V_2$, $C_1$ and $C_2$
\bqa
\label{eq:vert3}
%
%
AA~~:~~C_1 & = & K_1\nonumber \\
C_2 & = & K_2
\nonumber \\
\nonumber \\
%
%
AZ~~:~~C_1 & = & - \frac{c_w}{s_w} K_1 \nonumber \\
C_2 & = & - \frac{c_w}{s_w} K_2
\nonumber \\
\nonumber \\
%
%
ZZ~~:~~C_1 & = & \frac{c_w^2}{s_w^2} K_1\nonumber \\
C_2 & = & \frac{c_w^2}{s_w^2} K_2
\nonumber \\
\nonumber \\
%
%
W^-W^+~~:~~C_1 & = & \frac{1}{s_w^2} K_1\nonumber \\
C_2 & = & \frac{1}{s_w^2} K_2
\eqa
where
\bqa
K_1 &=& - \frac{1}{3} \lambda_{HV}+\frac{3}{4}(1-\xi)
\nonumber \\
\nonumber \\
K_2 &=& p^2\left(\frac{21\xi-17}{24}+\frac{\lambda_{HV}}{3}\right) 
- \xi \frac{\xi+3}{4} m_W^2 
\eqa

\vspace{0.3cm}
\leftline{{\bf Fermion-Fermion effective vertices}}
\vspace{0.3cm}

\noindent The generic effective vertex is
\bqa
{\rm Vert}(f_1,f_2) = \frac{ie^2}{\pi^2}
\left[\left(C_- \Omega^- + C_+\Omega^+\right)\rlap/p_1 + C_0
\right]
\eqa
with ${\rm Vert}(f_1,f_2)$ given in fig.~\ref{fig:1} $(d)$ and 
with the actual values of $f_1$, $f_2$, $C_-$, $C_+$ and $C_0$
\bqa
\label{eq:vert4}
%
%
u u ~~:~~C_- & = & \frac{Q_u^2}{c_w^2} 
\left(\frac{\lambda_{HV}}{16}-\frac{1-\xi}{24}  \right)\nonumber\\ 
C_+ & = & \left(\frac{I_{3u}^2}{s_w^2 c_w^2}
- \frac{ 2 Q_{u} I_{3u}}{c_w^2} + \frac{Q_u^2}{c_w^2}
+ \frac{1}{2 s_w^2} 
\left(V_{u d} V_{d u}^\dagger\right)
 \right) 
\left(\frac{\lambda_{HV}}{16}-\frac{1-\xi}{24}  \right)
\nonumber \\
C_0 & = & \frac{m_{u} Q_{u} }{8 c_w^2} \left( Q_{u} - I_{3u} \right) 
\left( \lambda_{HV}-\frac{1-\xi}{4}  \right)
\nonumber \\
\nonumber \\
%
%
d d ~~:~~C_- & = & \frac{Q_{d}^2}{c_w^2}
\left(\frac{\lambda_{HV}}{16}-\frac{1-\xi}{24}  \right) \nonumber \\
C_+ & = & \left(
\frac{I_{3{d}}^2}{s_w^2 c_w^2}
- \frac{ 2 Q_{d} I_{3d}}{c_w^2} + \frac{Q_{d}^2}{c_w^2}
+ \frac{1}{2 s_w^2}
\left(V_{u d} V_{d u}^\dagger\right) 
\right) 
\left(\frac{\lambda_{HV}}{16}-\frac{1-\xi}{24}  \right)
\nonumber \\
C_0 & = & \frac{m_{d} Q_{d}}{8 c_w^2} \left( Q_{d} - I_{3d}  \right) 
\left( \lambda_{HV}-\frac{1-\xi}{4}  \right)
\nonumber \\
\nonumber \\
%
%
l l ~~:~~C_- & = &  \frac{Q_{l}^2}{c_w^2}
\left(\frac{\lambda_{HV}}{16}-\frac{1-\xi}{24}  \right)  \nonumber \\
C_+ & = &  \left(\frac{I_{3l}^2}{s_w^2 c_w^2}
- \frac{ 2 Q_{l} I_{3l}}{c_w^2} + \frac{Q_{l}^2}{c_w^2}
+ \frac{1}{2 s_w^2} \right) 
\left(\frac{\lambda_{HV}}{16}-\frac{1-\xi}{24}  \right)
\nonumber \\
C_0 & = & \frac{m_{l}Q_{l}}{8 c_w^2} \left(  Q_{l} - I_{3l} \right) 
\left( \lambda_{HV}-\frac{1-\xi}{4}  \right)
\nonumber \\
\nonumber \\
%
%
\nu_l \nu_l ~~:~~C_- & = &  0 \nonumber \\
C_+ & = &  \frac{1}{s_w^2} \left(\frac{I^2_{3\nu_l}}{c_w^2}
+ \frac{1}{2} \right) \left(\frac{\lambda_{HV}}{16}-\frac{1-\xi}{24}  \right)
 \nonumber \\
C_0 & = & 0
\eqa

\subsubsection{Bosonic contribution to the vertices with 3 legs}

The generic 3-point vertices appearing in our calculation  are drawn in Fig.~\ref{fig:2}. As already pointed out, 
we limit ourselves to list the non vanishing cases, while 
the full set of results is available in~\cite{formfiles}. We found 43 non zero ${\rm R_2}$ 
vertices in the $R_\xi$ gauge, classified  in Table~\ref{tab:tab1}.
\begin{table}
\begin{center}
{\small Scalar-Scalar-Scalar vertices:}
\vspace{0.4cm}

\begin{tabular}{lll}
${\rm Vert}(H,H,H)$,   &  ${\rm Vert}(H,\chi,\chi)$,  &  ${\rm Vert}(H,\phi^+,\phi^-)$. 
\end{tabular}
\end{center}
\begin{center}
{\small Vector-Scalar-Scalar vertices:}
\vspace{0.4cm}

\begin{tabular}{llll}
${\rm Vert}(A,H,\chi)$,    & ${\rm Vert}(A,\phi^+,\phi^-)$,  & ${\rm Vert}(Z,H,\chi)$,  & ${\rm Vert}(Z,\phi^+,\phi^-)$,  \\
${\rm Vert}(W^-,H,\phi^+)$,    & ${\rm Vert}(W^-,\chi,\phi^+)$,  & 
${\rm Vert}(W^+,H,\phi^-)$,  & ${\rm Vert}(W^+,\chi,\phi^-)$.  
\end{tabular}
\end{center}
\begin{center}
{\small Scalar-Vector-Vector vertices:}
\vspace{0.4cm}

\begin{tabular}{llll}
${\rm Vert}(H,A,A)$,   & ${\rm Vert}(H,A,Z)$,  & ${\rm Vert}(H,Z,Z)$,  
                       & ${\rm Vert}(H,W^+,W^-)$, \\ 
${\rm Vert}(\phi^-,A,W^+)$,   & ${\rm Vert}(\phi^+,A,W^-)$ & ${\rm Vert}(\phi^-,Z,W^+)$, & ${\rm Vert}(\phi^+,Z,W^-)$.
\end{tabular}
\end{center}
\begin{center}
{\small Vector-Vector-Vector vertices:}
\vspace{0.4cm}

\begin{tabular}{ll}
${\rm Vert}(A,W^+,W^-)$,   & ${\rm Vert}(Z,W^+,W^-)$.
\end{tabular}
\end{center}
\begin{center}
{\small Scalar-Fermion-Fermion vertices:}
\vspace{0.4cm}

\begin{tabular}{llll}
${\rm Vert}(H,u,u)$,  & ${\rm Vert}(H,d,d)$, & ${\rm Vert}(H,l,l)$, 
                   &     \\ 
${\rm Vert}(\chi,u,u)$, & ${\rm Vert}(\chi,d,d)$,  & ${\rm Vert}(\chi,l,l)$, 
&   \\ 
${\rm Vert}(\phi^-,d,u)$, & ${\rm Vert}(\phi^-,l,\nu_l)$, & ${\rm Vert}(\phi^+,u,d)$,  & ${\rm Vert}(\phi^+,\nu_l,l)$.   
\end{tabular}
\end{center}
\begin{center}
{\small Vector-Fermion-Fermion vertices:}
\vspace{0.4cm}

\begin{tabular}{llll}
${\rm Vert}(A,u,u)$, & ${\rm Vert}(A,d,d)$, & ${\rm Vert}(A,\nu_l,\nu_l)$, 
                     & ${\rm Vert}(A,l,l)$, \\ 
${\rm Vert}(Z,u,u)$, & ${\rm Vert}(Z,d,d)$, & ${\rm Vert}(Z,\nu_l,\nu_l)$, 
                     & ${\rm Vert}(Z,l,l)$, \\ 

${\rm Vert}(W^-,d,u)$, & ${\rm Vert}(W^-,l,\nu_l)$, & ${\rm Vert}(W^+,u,d)$, 
                    & ${\rm Vert}(W^+,\nu_l,l)$.
\end{tabular}
\end{center}
\begin{center}
\caption{\label{tab:tab1} {The 43 non zero 3-point effective vertices
in the $R_\xi$ gauge.
In the Unitary gauge there are 23 non vanishing vertices, namely the 22 
listed here that do not contain $\chi$ or $\phi^\pm$ fields plus
${\rm Vert}(H,\nu_l,\nu_l)$.}}
\end{center}
\end{table}
\subsubsection{Bosonic contribution to the vertices with 4 legs}
All non vanishing generic 4-point vertices that appear in our calculation 
are drawn in Fig.~\ref{fig:3}. The full set of results can be found in~\cite{formfiles}.
The 35 non zero ${\rm R_2}$ vertices in the $R_\xi$ gauge are classified in Table~\ref{tab:tab2}.

\begin{table}
\begin{center}
{\small Scalar-Scalar-Scalar-Scalar vertices:}
\vspace{0.4cm}

\begin{tabular}{lll}
${\rm Vert}(H,H,H,H)$,  & ${\rm Vert}(H,H,\chi,\chi)$,  & ${\rm Vert}(H,H,\phi^-,\phi^+)$,  \\
${\rm Vert}(\chi,\chi,\chi,\chi)$,  & ${\rm Vert}(\chi,\chi,\phi^-,\phi^+)$,  
                        & ${\rm Vert}(\phi^-,\phi^+,\phi^-,\phi^+)$.  
\end{tabular}
\end{center}
\begin{center}
{\small Scalar-Scalar-Vector-Vector effective vertices:}
\vspace{0.4cm}

\begin{tabular}{llll}
${\rm Vert}(H,H,A,A)$,$\!\!$ & ${\rm Vert}(H,H,A,Z)$,$\!\!$ & 
                ${\rm Vert}(H,H,Z,Z)$,$\!\!$ & ${\rm Vert}(H,H,W^-,W^+)$,                     $\!$\\ 
${\rm Vert}(H,\phi^+,W^-,A)$,$\!\!$ & ${\rm Vert}(H,\phi^+,W^-,Z)$,$\!\!$ &  
                ${\rm Vert}(\chi,\chi,A,A)$,$\!\!$ & ${\rm Vert}(\chi,\chi,A,Z)$,             $\!$\\ 
${\rm Vert}(\chi,\chi,Z,Z)$,$\!\!$ & ${\rm Vert}(\chi,\chi,W^-,W^+)$,$\!\!$ &
                ${\rm Vert}(\chi,\phi^+,W^-,A)$,$\!\!$ & ${\rm Vert}(\chi,\phi^+,W^-,Z)$,     $\!$\\ 
${\rm Vert}(\phi^-,H,A,W^+)$,$\!\!$ & ${\rm Vert}(\phi^-,H,Z,W^+)$,$\!\!$ & 
                ${\rm Vert}(\phi^-,\chi,A,W^+)$,$\!\!$ & ${\rm Vert}(\phi^-,\chi,Z,W^+)$,     $\!$\\ 
${\rm Vert}(\phi^-,\phi^+,A,A)$,$\!\!$ & ${\rm Vert}(\phi^-,\phi^+,A,Z)$, $\!\!$&
                ${\rm Vert}(\phi^-,\phi^+,Z,Z)$,$\!\!$ & ${\rm Vert}(\phi^-,\phi^+,W^-,W^+)$. $\!$ 
\end{tabular}

\end{center}
\begin{center}
{\small Vector-Vector-Vector-Vector effective vertices:}
\vspace{0.4cm}

\begin{tabular}{lll}
${\rm Vert}(A,A,A,A)$,  & ${\rm Vert}(A,A,A,Z)$,  & ${\rm Vert}(A,A,Z,Z)$,  \\
${\rm Vert}(A,Z,Z,Z)$,  & ${\rm Vert}(Z,Z,Z,Z)$,  & ${\rm Vert}(A,A,W^-,W^+)$,  \\
${\rm Vert}(A,Z,W^-,W^+)$,  & ${\rm Vert}(Z,Z,W^-,W^+)$,  & ${\rm Vert}(W^-,W^+,W^-,W^+)$.
\end{tabular}

\end{center}
\begin{center}
\caption{\label{tab:tab2} {The 35 non zero 4-point effective vertices
in the $R_\xi$ gauge. In the Unitary gauge there are 14  
non vanishing vertices, namely all those ones that do not contain 
$\chi$ or $\phi^\pm$ fields.}}
\end{center}
\end{table}

\subsection{The Unitary gauge}
We follow again the notations of Fig.~\ref{fig:1}.

\subsubsection{Bosonic contribution to the vertices with 2 legs}

\vspace{0.3cm}
\leftline{{\bf Scalar-Scalar effective vertices}}
\vspace{0.3cm}

\noindent The generic effective vertex is
\bqa
{\rm Vert}(S_1,S_2) = \frac{ie^2}{16 \pi^2 s_w^2} C
\eqa
with ${\rm Vert}(S_1,S_2)$ given in fig.~\ref{fig:1} $(a)$ and 
with the actual values of $S_1$, $S_2$ and $C$
\bqa
%
%
HH~~:~~C & = & 
 \frac{5}{6}\, p_1^2  \left(1+\frac{1}{2 c_w^2} \right)
-\frac{9}{40}\frac{p_1^4}{m_W^2}
- m_W^2 \left(1+\frac{1}{2 c_w^4} \right)
  \left(\frac{1}{4}+ 3 \lambda_{HV} \right)
\eqa

\vspace{0.3cm}
\leftline{{\bf Vector-Scalar effective vertices}}
\vspace{0.3cm}
\noindent No contribution is found in the Unitary gauge.

\vspace{0.3cm}
\leftline{{\bf Vector-Vector effective vertices}}
\vspace{0.3cm}

\noindent The generic effective vertex is
\bqa
{\rm Vert}(V_1,V_2) = \frac{ie^2}{8 \pi^2} 
\left(C_1 \,p_{1\alpha} p_{1\beta} + C_2 \,g_{\alpha \beta}\right)
\eqa
with ${\rm Vert}(V_1,V_2)$ given in fig.~\ref{fig:1} $(c)$ and 
with the actual values of $V_1$, $V_2$, $C_1$ and $C_2$
\bqa
%
%
AA~~:~~C_1 & = & K_1\nonumber \\
C_2 & = & K_2
\nonumber \\
\nonumber \\
%
%
AZ~~:~~C_1 & = & - \frac{c_w}{s_w} K_1 \nonumber \\
C_2 & = & - \frac{c_w}{s_w} K_2
\nonumber \\
\nonumber \\
%
%
ZZ~~:~~C_1 & = & \frac{c_w^2}{s_w^2} K_1\nonumber \\
C_2 & = & \frac{c_w^2}{s_w^2} K_2
\nonumber \\
\nonumber \\
%
%
W^-W^+~~:~~C_1 & = &  \frac{1}{s_w^2} K_3\nonumber \\
C_2 & = & \frac{1}{s_w^2} K_4
\eqa
where
\bqa
K_1 &=&  -\frac{1}{3} \left(\lambda_{HV}-5\right)-\frac{17}{60} \,\frac{p_1^2}{m_W^2}
\nonumber \\
\nonumber \\
K_2 &=& \frac{3}{4} m_W^2+\frac{1}{3}\,p_1^2\,\left(\lambda_{HV}-\frac{23}{4} 
                    \right)+\frac{37}{120}\,\,\frac{p_1^4}{m_W^2}
\nonumber \\
\nonumber \\
K_3 &=& -\frac{1}{3} \left(\lambda_{HV}-\frac{5}{2}
         -\frac{9}{8} c_w^2 \right)
        +\frac{11}{24} c_w^4-\frac{17}{120}\,\frac{p_1^2}{m_W^2}
         \left(1+ c_w^4  \right)  
\nonumber \\
\nonumber \\
K_4 &=&  \frac{3}{8} \frac{m_W^2}{c_w^2}\left(s_w^2+c_w^4+c_w^6 \right)
        + p_1^2\left[\frac{\lambda_{HV}}{3}-\frac{7}{8}
        -\frac{7}{16} c_w^2\left(1+\frac{29}{21}c_w^2 \right) \right] 
        \nonumber \\
        &&+\frac{37}{240}\frac{p_1^4}{m_W^2} \left(1+c_w^4 \right) 
\eqa

\vspace{0.3cm}
\leftline{{\bf Fermion-Fermion effective vertices}}
\vspace{0.3cm}

\noindent The generic effective vertex is
\bqa
{\rm Vert}(f_1,f_2) = \frac{ie^2}{\pi^2}
\left[\left(C_- \Omega^- + C_+\Omega^+\right)\rlap/p_1 + C_0
\right]
\eqa
with ${\rm Vert}(f_1,f_2)$ given in fig.~\ref{fig:1} $(d)$ and 
with the actual values of $f_1$, $f_2$, $C_-$, $C_+$ and $C_0$
\bqa
%
%
u u ~~:~~C_- & = & 
\frac{Q_u^2}{16 c_w^2} 
\left[ 
\lambda_{HV}
+\frac{s_w^2}{m_Z^2} \left(\frac{p_1^2}{4}-\frac{2}{3} m_Z^2
    -\frac{5}{6} m_u^2 \right)
\right]
\nonumber\\ 
C_+ & = & \frac{\lambda_{HV}}{16} \left[\frac{I_{3u}^2}{s_w^2 c_w^2}
- \frac{ 2 Q_{u} I_{3u}}{c_w^2} + \frac{Q_u^2}{c_w^2}
+ \frac{1}{2 s_w^2} 
\left(V_{u d} V_{d u}^\dagger\right)
 \right] 
 \nonumber \\
&&+ \frac{s_w^2}{16 m_Z^2 c_w^2}
    \left(\frac{p_1^2}{4}-\frac{2}{3} m_Z^2
    -\frac{5}{6} m_u^2 \right)
          \left(Q_u-\frac{I_{3u}}{s_w^2} \right)^2
 \nonumber \\
&& + \frac{V_{u d} V_{d u}^\dagger}{32 m_W^2 s_w^2}  
    \left(\frac{p_1^2}{4}-\frac{2}{3} m_W^2
    -\frac{5}{6} m_d^2 \right)
\nonumber \\
C_0 & = &  \frac{Q_u m_u}{8 c_w^2} \left[                  
         \lambda_{HV}  \left(Q_u-I_{3u} \right)            
         +\frac{s_W^2}{4 m_Z^2}                            
          \left(Q_u-\frac{I_{3u}}{s_w^2} \right)           
          \left(\frac{p_1^2}{3}-m_Z^2-m_u^2 \right) \right]
\nonumber \\
\nonumber \\
%
%
d d ~~:~~C_- & = & 
\frac{Q_d^2}{16 c_w^2} 
\left[ 
\lambda_{HV}
+\frac{s_w^2}{m_Z^2} \left(\frac{p_1^2}{4}-\frac{2}{3} m_Z^2
    -\frac{5}{6} m_d^2 \right)
\right]
\nonumber \\
C_+ & = & \frac{\lambda_{HV}}{16} \left[\frac{I_{3d}^2}{s_w^2 c_w^2}
- \frac{ 2 Q_{d} I_{3d}}{c_w^2} + \frac{Q_d^2}{c_w^2}
+ \frac{1}{2 s_w^2} 
\left(V_{u d} V_{d u}^\dagger\right)
 \right] 
 \nonumber \\
&&+ \frac{s_w^2}{16 m_Z^2 c_w^2}
    \left(\frac{p_1^2}{4}-\frac{2}{3} m_Z^2
    -\frac{5}{6} m_d^2 \right)
          \left(Q_d-\frac{I_{3d}}{s_w^2} \right)^2
 \nonumber \\
&& + \frac{V_{u d} V_{d u}^\dagger}{32 m_W^2 s_w^2}  
    \left(\frac{p_1^2}{4}-\frac{2}{3} m_W^2
    -\frac{5}{6} m_u^2 \right)
\nonumber \\
C_0 & = &   \frac{Q_d m_d}{8 c_w^2} \left[                  
          \lambda_{HV}  \left(Q_d-I_{3d} \right)            
          +\frac{s_W^2}{4 m_Z^2}                            
           \left(Q_d-\frac{I_{3d}}{s_w^2} \right)           
           \left(\frac{p_1^2}{3}-m_Z^2-m_d^2 \right) \right]
\nonumber \\
\nonumber \\
%
%
l l ~~:~~C_- & = &  
\frac{Q_l^2}{16 c_w^2} 
\left[ 
\lambda_{HV}
+\frac{s_w^2}{m_Z^2} \left(\frac{p_1^2}{4}-\frac{2}{3} m_Z^2
    -\frac{5}{6} m_l^2 \right)
\right]
\nonumber \\
C_+ & = & \frac{\lambda_{HV}}{16} \left[\frac{I_{3l}^2}{s_w^2 c_w^2}
- \frac{ 2 Q_{l} I_{3l}}{c_w^2} + \frac{Q_l^2}{c_w^2}
+ \frac{1}{2 s_w^2} \right] 
 \nonumber \\
&&+ \frac{s_w^2}{16 m_Z^2 c_w^2}
    \left(\frac{p_1^2}{4}-\frac{2}{3} m_Z^2
    -\frac{5}{6} m_l^2 \right)
          \left(Q_l-\frac{I_{3l}}{s_w^2} \right)^2
 \nonumber \\
&& + \frac{1}{32 m_W^2 s_w^2}  
    \left(\frac{p_1^2}{4}-\frac{2}{3} m_W^2\right)
\nonumber \\
C_0 & = & \frac{Q_l m_l}{8 c_w^2} \left[                  
         \lambda_{HV}  \left(Q_l-I_{3l} \right)            
         +\frac{s_W^2}{4 m_Z^2}                            
          \left(Q_l-\frac{I_{3l}}{s_w^2} \right)           
          \left(\frac{p_1^2}{3}-m_Z^2-m_l^2 \right) \right]
\nonumber \\
\nonumber \\
%
%
\nu_l \nu_l ~~:~~C_- & = &  0 \nonumber \\
C_+ & = &  
\frac{\lambda_{HV}}{16 s_w^2} 
 \left(\frac{1}{2}+\frac{I^2_{3\nu_l}}{c_w^2} \right) 
+ \frac{I^2_{3\nu_l}}{16 m_Z^2 c_w^2 s_w^2}
    \left(\frac{p_1^2}{4}-\frac{2}{3} m_Z^2\right)
 \nonumber \\
&& + \frac{1}{32 m_W^2 s_w^2}  
    \left(\frac{p_1^2}{4}-\frac{2}{3} m_W^2 
    -\frac{5}{6} m_l^2 \right)
\nonumber \\
C_0 & = & 0
\eqa

\subsubsection{Bosonic contribution to the vertices with 3 legs}
The generic 3-point vertices appearing in our calculation are drawn in Fig.~\ref{fig:2}. As before
the full set of results is available in~\cite{formfiles}. 
We found 23 non zero ${\rm R_2}$  vertices in the Unitary gauge, classified  in Table~\ref{tab:tab1}.

\subsubsection{Bosonic contribution to the vertices with 4 legs}
All non vanishing generic 4-point vertices that appear in our calculation 
are drawn in Fig.~\ref{fig:3}. The full set of results is presented in~\cite{formfiles}.
The 14 non zero ${\rm R_2}$ vertices in the Unitary gauge are classified in Table~\ref{tab:tab2}.
\section{\label{checks}Checks}
All our formulae have been obtained cross-checking two independent calculations.
To further check our results, we used the fact that the ${\rm R}= {\rm R_1}+{\rm R_2}$ contribution to physical quantities 
should be independent of the chosen gauge.
In particular, parametrizing the gauge boson self-energies as follows
\bqa
\Sigma_V^{\mu\nu}(p) &=& g^{\mu\nu}\, \Sigma_{V0}(p^2) +p^\mu p^\nu\,\Sigma_{V1}(p^2)~~~~{\rm with}~~~{V= Z,W,\gamma}\,,
\eqa
we verified that the ${\rm R}$ contribution to 
$\Sigma_{W0}(M_W^2)$, $\Sigma_{Z0}(M_Z^2)$ and $\Sigma_{\gamma 0}(0)$ is the same in both the $R_\xi$ and the Unitary gauge. 
In addition, in the case of both gauges, we checked all of 
the 2-point like Ward Identities presented in~\cite{ewrational} involving 
${\rm Vert}(S_1,S_2)$, ${\rm Vert}(V,S)$ and ${\rm Vert}(V_1,V_2)$.  

To test the 3-point sector, we computed the ${\rm R}= {\rm R_1}+{\rm R_2}$ contribution to the process
$
H \to \gamma \gamma\,.
$
Again, we found the same answer working in both gauges, obtaining an expression 
for ${\rm R}$ in full agreement with that one presented in~\cite{Bardin:1999ak}.
As for the 4-point sector, we checked that, in the limit $\xi \to 1$, we fully reproduce the effective vertices
presented in~\cite{ewrational}.

Finally, in the case of the $R_\xi$ gauge, we computed ${\rm R_2}$ using
both the following two equivalent representations for the massive 
gauge boson propagators
\bqa
&& {-i} \left(\frac{g_{\alpha\beta}}{p^2-M_V^2}
  -(1-\xi)\frac{p_\alpha p_\beta}{(p^2-M_V^2)(p^2-\xi M_V^2)}\right) 
\,~~{\rm and}~~\nl
&& {-i} \left(\frac{g_{\alpha\beta}}{p^2-M_V^2}
        -\frac{p_\alpha p_\beta} {M_V^2(p^2-M_V^2)}
        +\frac{p_\alpha p_\beta} {M_V^2(p^2-\xi M_V^2)} \right)\,,
\eqa
always finding the same results. Since the two expressions lead 
to different integrals in the intermediate stages of the calculation, 
this provides a strong consistency check of our procedure.

As a last remark notice that, when working in the Unitary gauge, we take the 
limit $\xi \to \infty$ {\em before} integrating over the loop momentum. The fact that this gives the same result for ${\rm R}$ as in a generic $R_\xi$ gauge 
in the above mentioned cases 
{\em provided the same prescription  is used in the calculation of ${\rm R_1}$} is an explicit check of the equivalence
of the limits $\xi \to \infty$ after or before the loop momentum integration in the definition of the Unitary gauge at 1-loop. 

\section{Conclusions}
We presented the full set of Feynman rules producing the rational terms of kind ${\rm R_2}$ needed to perform 
any 1-loop calculation in the Electroweak Standard Model in the $R_{\xi}$ gauge and in the Unitary gauge.
In a few physical cases we also checked the independence of the full rational piece
${\rm R}= {\rm R_1}+{\rm R_2}$ of the chosen gauge and, in the case 
of the Unitary gauge, of the order between the limit $\xi \to \infty$
and the integration over the loop momentum. 
Our results can be used to transform tree level packages based on gauges 
other that the 't Hooft-Feynman one into 1-loop calculators with the help of the 
OPP or Generalized Unitarity techniques.

\section*{Acknowledgments}
R.P.'s and I.M.'s research was partially supported by the RTN
European Programme MRTN-CT-2006-035505 (HEPTOOLS, Tools and Precision
Calculations for Physics Discoveries at Colliders).
 M.V.G.'s research was supported by INFN.
The research of R.P. and M.V.G. was also supported 
by the MEC project FPA2008-02984.
R.P. also acknowledges the financial support of the bilateral
INFN/MICINN program ACI2009-1045 (Aspects of Higgs physics at the LHC).

\end{document}